\documentclass[english,5p, twocolumn]{elsarticle}
\usepackage[latin9]{inputenc}
\usepackage{graphicx}
\usepackage{fancyhdr}

\makeatletter

\providecommand{\tabularnewline}{\\}

\@ifundefined{textcolor}{}
{%
 \definecolor{BLACK}{gray}{0}
 \definecolor{WHITE}{gray}{1}
 \definecolor{RED}{rgb}{1,0,0}
 \definecolor{GREEN}{rgb}{0,1,0}
 \definecolor{BLUE}{rgb}{0,0,1}
 \definecolor{CYAN}{cmyk}{1,0,0,0}
 \definecolor{MAGENTA}{cmyk}{0,1,0,0}
 \definecolor{YELLOW}{cmyk}{0,0,1,0}
 }

\@ifundefined{showcaptionsetup}{}{%
 \PassOptionsToPackage{caption=false}{subfig}}
\usepackage{subfig}
\makeatother

\usepackage{babel}

\begin{document}

\title{Strong Coupling Optimization With Planar Spiral Resonators}

\author[rihu]{Avraham Klein}
\ead{avraham.klein@mail.huji.ac.il}

\author[rihu]{Nadav Katz\corref{cor1}}
\ead{katzn@phys.huji.ac.il}
\address[rihu]{Racah Institute, Hebrew University of Jerusalem, Givat Ram, Israel}

\cortext[cor1]{Corresponding Author. Telephone: 972-2-6586745, fax: 972-2-6586347}

\begin{abstract}
Planar spirals offer a highly scalable geometry appropriate for wireless
power trasfer via strongly coupled inductive resonators. We numerically
derive a set of geometric scale and material independent coupling
terms, and analyze a simple model to identify design considerations
for a variety of different materials. We use our model to fabricate
integrated planar resonators of handheld sizes, and optimize them
to achieve high Q factors, comparable to much larger systems, and
strong coupling over significant distances with approximately constant
efficiency.
\end{abstract}
\maketitle
\section{Introduction}

Wireless power transfer via resonant magnetic coupling has attracted
considerable attention in recent years. This is due both to its elegance
and to its possible applicability at many different size scales, from
powering spaceships and cars \citep{Sedwick2010,Imura2009}, and down
to handheld scale devices \citep{FredySegura-Quijano2008} and microdevice
coupling \citep{Ghovanloo2007}. Such coupling depends strongly on
two predominantly geometric properties: The devices involved must
be high quality resonators, and they must have far-reaching magnetic
fields \citep{Haus1984,Karalis2008}. Thus, the geometric design of
the resonators is of utmost importance. Moreover, the geometry of
a device is inherently size independent, and this means that if a
design exists that can be built at different size scales, using different
materials, then the same considerations will apply to all variations.
The planar spiral is such a design, being both simple and quasi two-dimensional.
Thus, for example, planar spiral designs can easily be etched on a
thin substrate and incorporated into current handheld devices. 

In this Letter, we analyze a planar spiral model and numerically computed
coupling terms in order to identify optimal design considerations
for different materials at a desired size scale. For example, we show
how high $T_{C}$ superconductors can be designed so as to achieve
very strong coupling. We use our method to optimize the design of
a device similar in size to currently used handheld devices, using
inexpensive materials, by identifying the properties of the dominant
dielectric loss channel for the size/materials involved and using
capacitative loading to compensate. We achieve very high Q factors
that are typical of much larger devices and strong coupling over significant
ranges. We then show that the coupling between the devices is robust,
being almost constant over the entire coupling range.

\section{Theoretical Model}

Coupling between two high-Q resonators is adequately described by
Coupled-Mode Theory \citep{Haus1984,Kurs2007}. A source and destination
device can be represented by complex-valued variables $a_{1},a_{2}$
normalized so that $\left|a_{n}^{2}\right|$ is the energy in a resonator,
obeying the relation:\begin{eqnarray}
-i\omega a_{1} & = & -\left[i\omega_{0}+\Gamma_{1}\right]a_{1}+i\kappa a_{2}+F\label{eq:CMTbasic}\\
-i\omega a_{2} & = & -\left[i\omega_{0}+\Gamma_{2}\right]a_{2}+i\kappa a_{1}\label{eq:CMTbasic2}\end{eqnarray}
where $\Gamma_{m}=\left(1+k_{m}\right)\gamma_{m}$ is the loaded dissipation
factor of the device, $k_{m}$ is a coupling coefficient to some load
or measuring device and $\gamma_{m}=\frac{\omega_{0}}{2Q_{m}}$ is
the unloaded dissipation factor. $\kappa$ represents the coupling
between the two devices and $F$ is a forcing term for the source
$a_{1}$. Solving eqs. \ref{eq:CMTbasic}-\ref{eq:CMTbasic2} yields \emph{frequency splitting}:
$\left|a_{2}\right|^{2}$ is maximized and the devices transfer energy
efficiently when $\omega-\omega_{0}=\pm\sqrt{\kappa^{2}-\left(\Gamma_{1}^{2}+\Gamma_{2}^{2}\right)/2}$.
The regime where this splitting takes place is called the \emph{strong
coupling} regime. For identical resonators $\left(k_{1}=k_{2}=k_{c},\gamma_{1}=\gamma_{2}=\gamma\right)$
this splitting is possible when:

\begin{equation}
\frac{\kappa}{\left(1+k_{c}\right)\gamma}\geq1\label{eq:strongCpl}\end{equation}
Thus a Figure of Merit for such a system is the quantity $\kappa/\gamma=Qk$
where $k$ is a dimensionless coupling term. In this we assume $k_c \leq 1$ so as to maintain the strong coupling, or equivalently, keep the loaded quality factor of the devices high. For magnetically coupled
systems $k=M/L$, $M,L$ being the mutual and self inductances of
the devices respectively. The efficiency in this regime for identical
resonators at the split frequencies is constant and obeys:

\begin{eqnarray}
\eta & = & \frac{k_c}{2\left(1+k_c\right)}\label{eq:eff}\end{eqnarray}
The upper limit of $\eta\leq1/4$ (for $k_c \leq 1$) appears because the devices are
identical. Higher efficiency is obtained with proper loading \citep{Kurs2007,Karalis2008}.

\begin{figure}
\noindent \includegraphics[bb=0bp 55bp 960bp 660bp,clip,width=1\columnwidth]{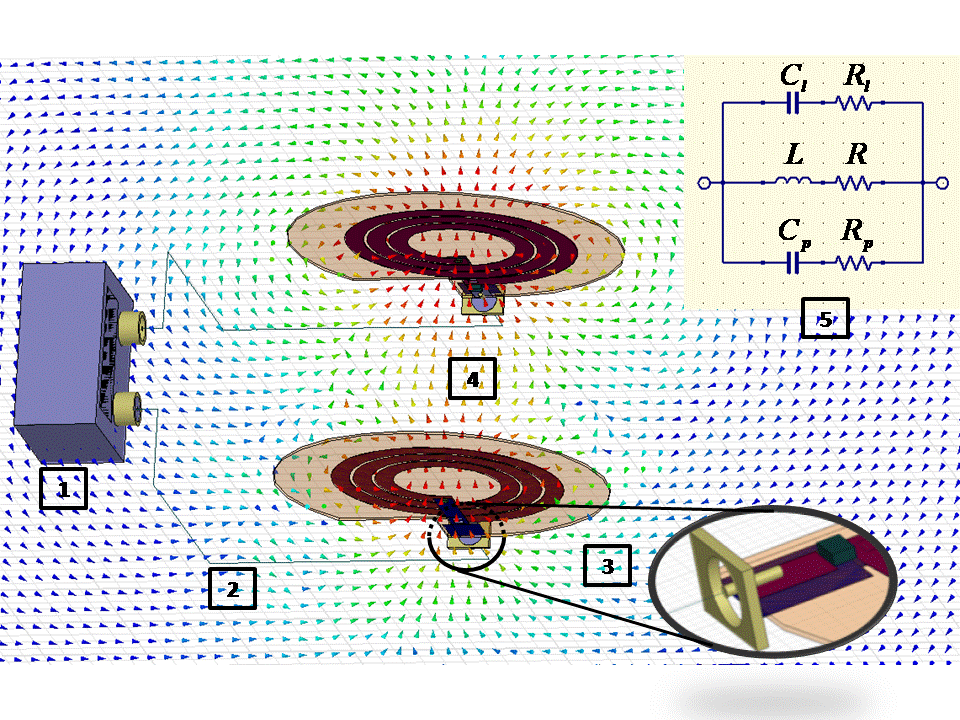}

\caption{Schematic of the experimental setup. A network analyzer (1) is connected
via coaxial feedlines (2) to a pair of coplanar coupled spirals (the spiral geometry is detailed 
in Fig. \ref{Flo:spiralSketch}). Each
spiral is matched to the feedlines via a bonded chip capacitor (3).
The S-parameters can be analyzed to obtain the Q factor, coupling
and power transfer efficiency between devices. The figure also shows
a numerical simulation of the magnetic coupling fields (4), computed
using HFSS 11. (5) shows a simplified lumped-element circuit that
we have used in our theoretical model.}
\label{Flo:Setup}
\end{figure}

\begin{figure}
\subfloat[Top-down view, with substrate layer hidden.]{\begin{centering}
\includegraphics[width=0.9\columnwidth]{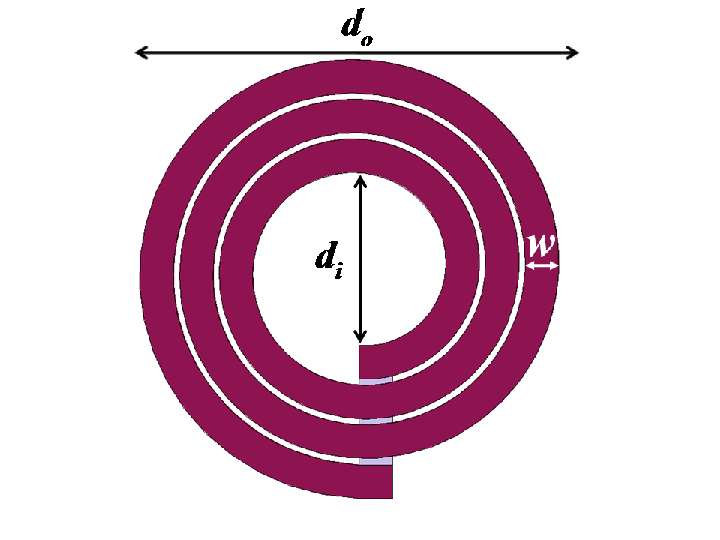}\label{Flo:SpiralTD}
\par\end{centering}
}

\subfloat[Side view, with substrate layer.]{\begin{centering}
\includegraphics[bb=0bp 200bp 720bp 400bp,clip,width=0.9\columnwidth]{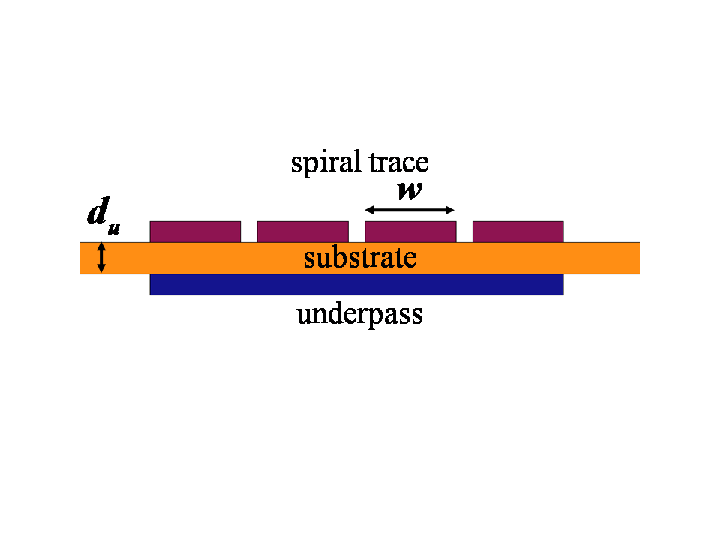}\label{Flo:SpiralSV}
\par\end{centering}
}
\caption{Sketch of a basic spiral design, and the spiral parameters referred to in our theoretical model.}
\label{Flo:spiralSketch}
\end{figure}

\begin{figure}
\subfloat[\emph{Upper right}:\emph{ }The coupling coefficient table $k\left(n,w,d=1.8d_{o}\right)$
used in the analysis in this work. \emph{Lower left}: $QM/L$ for
a copper spiral on a Rogers 4350B substrate. Ohmic and dielectric
losses dominate.]{\begin{centering}
\includegraphics[width=0.9\columnwidth]{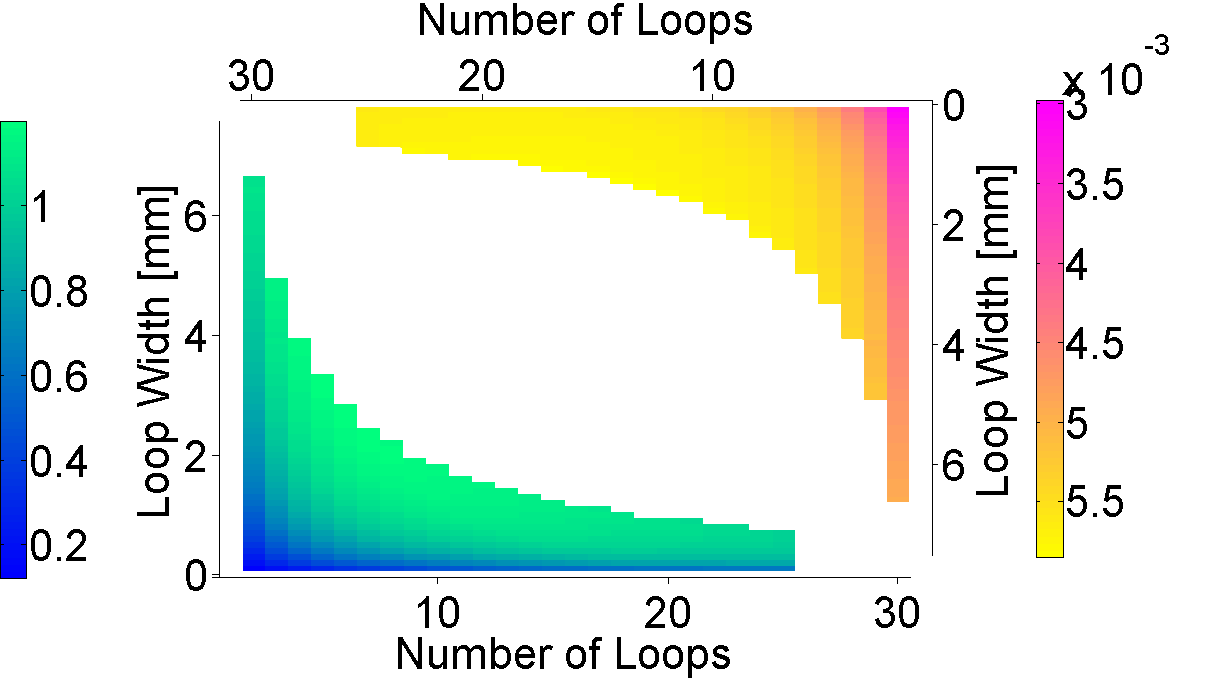}\label{Flo:kandrogers}
\par\end{centering}

}

\subfloat[\emph{Upper right}:\emph{ }$QM/L$ for a copper spiral on a sapphire
substrate. Ohmic losses dominate, and removing the dielectric loss
allows a very clear preferred design. \emph{Lower left}: $QM/L$ for
a superconducting YBCO film on sapphire (assumed conductivity: $5\times10^{3}\sigma_{Cu}$
\citep{Lancaster2006}). Metallic losses dominate but the ohmic loss
reduction yields two orders of magnitude better performance.]{\begin{centering}
\includegraphics[width=0.9\columnwidth]{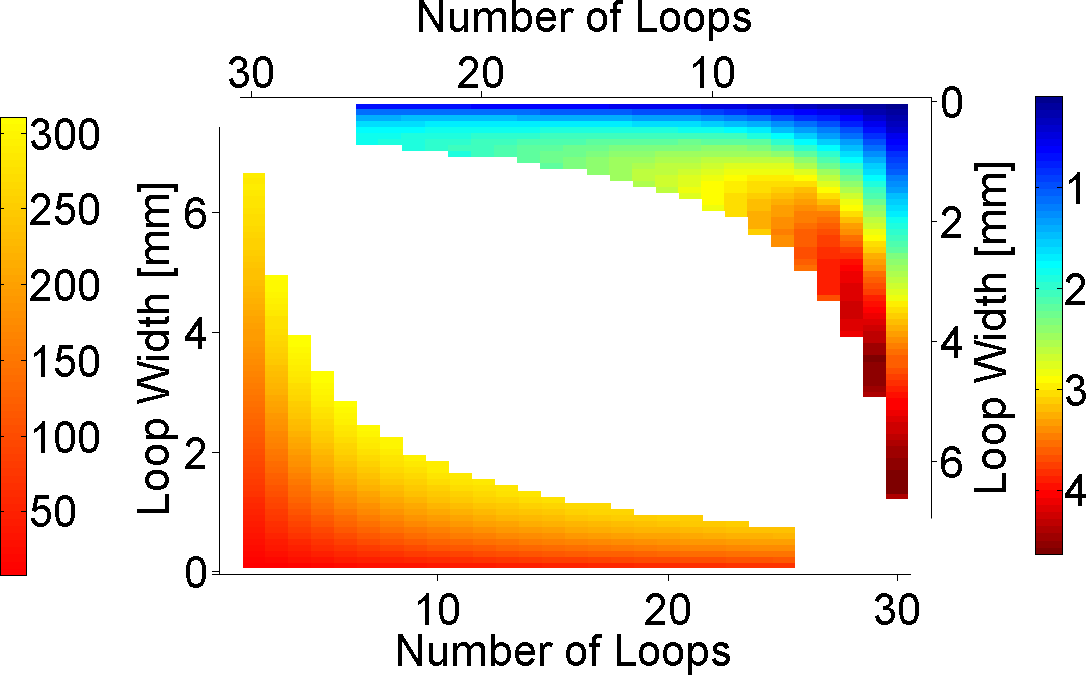}\label{Flo:sapphireandhts}
\par\end{centering}

}

\caption{$QM/L$ optimization for various designs, showing that reducing dielectric
and ohmic losses changes design considerations and quality. The model
dimensions used are the same as that of Device \#1 detailed later.}

\label{Flo:designLoss}
\end{figure}

Such coupling can be achieved by using planar spiral resonators driven
at quasi-static frequencies. It is well known that such resonators can be modeled as
lumped RLC resonant circuits. For a description of the considerations involved in modeling 
spirals and related planar spirals, we refer the reader to Refs. \citep{Bilotti2007a,Bilotti2007}.
Fig. \ref{Flo:Setup} shows the schematic
of such a setup and a simple equivalent circuit for a spiral. Such
a resonator is described by inner and outer diameters $d_{i},d_{o}$,
number of loops $n$ and loop width $w$, and has a resonant frequency
$\omega_{0}$. Additionally, an underpass strip usually connects the
spiral inner and outer extremities, at a separation $d_{u}$. Fig. \ref{Flo:spiralSketch} shows 
a sketch of a basic spiral. Effective
$L$ values can be found using current sheet approximations \citep{Mohan1999}.
The capacitance can be written as $C=C_{l}+C_{p}$, where $C_{l},C_{p}$
are the inter-coil and coil-underpass capacitances, approximated with
coplanar waveguide \citep{K.C.Gupta1996,Pieters2001} (with loops
coupling in series) and parallel-plate \citep{Gao2006} formulas.
The coupling coefficient $k=M/L$ can be extracted numerically. In
this work we constructed a table of $k\left(n,w,d\right)$ values
($d$ being the coplanar distance between spirals) using the \emph{Fasthenry}
multipole expansion tool \citep{Kamon1994} (see fig. \ref{Flo:kandrogers}).
These values are scale/material independent so that the same table
predicts behaviour and design parameters for a wide variety of possible
designs. Finally the metallic losses consist of radiative and ohmic
losses:

\begin{eqnarray}
R & = & \frac{1}{\sigma\xi}\times\frac{2\pi\sum r_{i}}{2w\cdot\delta_{s}}+\label{eq:losses}\\
 &  & \sqrt{\frac{\mu_{0}}{\epsilon_{0}}}\left[\frac{\pi}{6}\left(\frac{\omega}{c}\right)^{4}\left(\sum_{i=1}^{n}r_{i}^{2}\right)^{2}+\frac{4n^{2}}{3\pi^{3}}\left(\frac{\omega}{c}\right)^{2}d_{u}^{2}\right]\nonumber \end{eqnarray}
where the first term represents ohmic losses: $\sigma,\delta_{s}$
are the trace conductivity and skin depth, $\xi$ is an empirical
current-crowding factor\citep{Peck1995}, and $\sum r_{i}$ is an approximation
to the spiral length as a sum of concentric circles with respective
radii $r_{i}$. The second term describes magnetic and electric dipole
resistance. The resistances of the capacitative channels $R_{p},R_{l}$
are determined by the substrate loss tangent $\tan\delta$. The quality
factor of the spiral is then:

\begin{equation}
Q=\left(\tan\delta+\frac{R}{\omega_{0}L}\right)^{-1}\label{eq:Q}\end{equation}

Examining eq. \ref{eq:Q} leads to a number of conclusions. First,
the only scale-dependent factor influencing $QM/L$ is the ohmic loss
channel in $R$. In all other terms (radiation loss, inductance and
capacitance) the scale dependencies cancel out. Next, $\tan\delta$
is independent of both scale \emph{and }geometry. Therefore when dielectric
loss dominates one cannot optimize the design. However, adding a high
quality capacitance $C_{ext}$ in parallel will both reduce the effective
loss, i.e. $\tan\delta\rightarrow\tan\delta\frac{C}{C+C_{ext}}$,
and allow optimization reducing other losses to be significant. Specifically,
reduction of loop number and wide traces are crucial, as they lead
to a reduction of the term $\sum r_{i}/w$. Finally, if the ohmic
losses can be reduced to the order of radiative losses, for example
by using superconductors, then a single optimal design can be adapted
to many different scales. Fig. \ref{Flo:designLoss} shows actual
design scans confirming these conclusions.

\begin{figure}
\includegraphics[width=0.9\columnwidth]{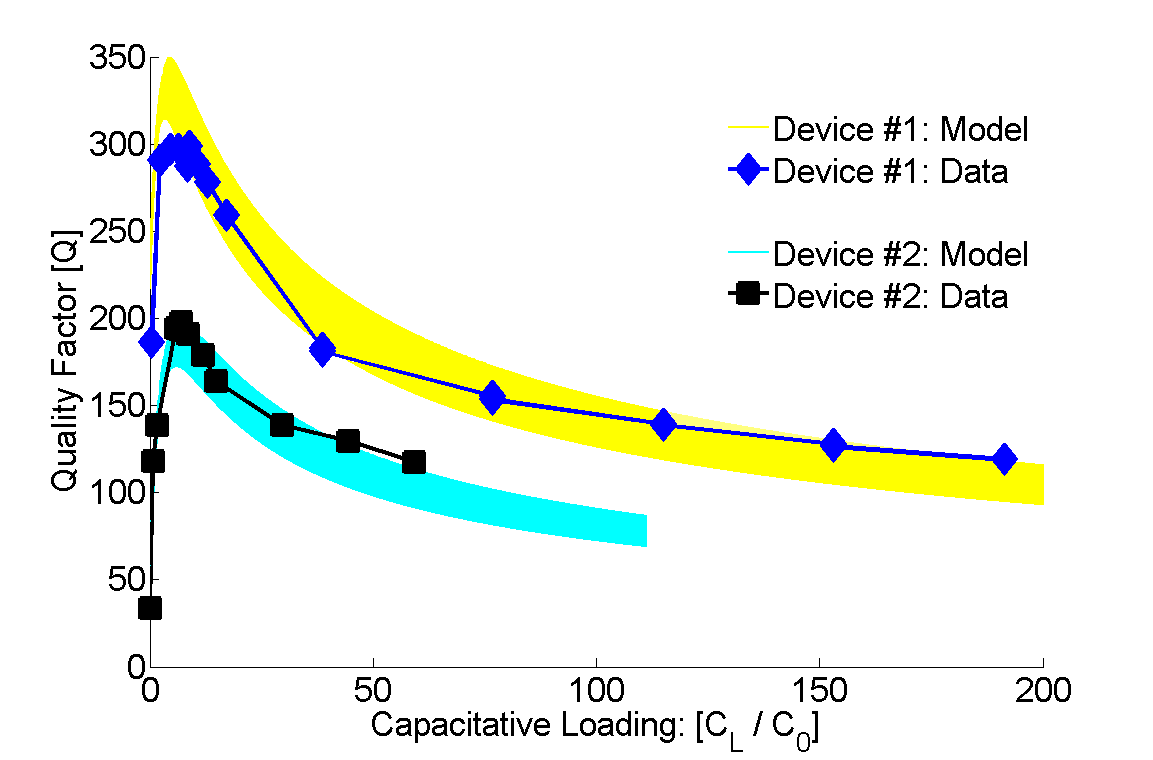}

\caption{Capacitative optimization of individual resonators. Note the sharp
peak at $C_{L}/C_{0}\sim7$.}
\label{Flo:capOpt}
\end{figure}

\section{Experimental Results and Discussion}

Two devices were built to test the coupling efficiency and design
optimization. Device dimensions were $d_{o}=60\, mm.,n=3,w=4.4\, mm.$
and $d_{i}/d_{o}=0.33$ (considered optimal\citep{Ghovanloo2007}).
Device \#1 was fabricated on a $0.51\, mm.$ thick Rogers 4350B substrate
($\tan\delta=0.0037$).{\scriptsize{} }Device \#2 was fabricated on
$1.55\, mm.$ thick FR4 ($\tan\delta\simeq0.0180$). The devices were
matched to an Agilent N5230A network analyzer as in fig. \ref{Flo:Setup}.
High-Q RF chip capacitors ($Q>1000$) were soldered in series for
network matching and in parallel for design optimization.

Fig. \ref{Flo:capOpt} shows the Q factor of devices as a function
of loading, showing that for optimal loading a sharp peak is obtained,
as predicted by eq. \ref{eq:Q}, with added capacitance affecting
the $\tan\delta,\omega_{0},R$ terms. $Q\simeq300$ for Device \#1
at the maximum, which is comparable to Q factors of air cored designs
an order of magnitude larger \citep{Sample2010,FredySegura-Quijano2008,Kurs2007}.
The theoretical area shown is for $3.5\leq\xi\leq4.6$, so using $\xi=4$
should give accurate results, and this is the number used in all theoretical
simulations in this work.

\begin{figure}
\includegraphics[bb=10bp 220bp 1130bp 655bp,clip,width=1\columnwidth]{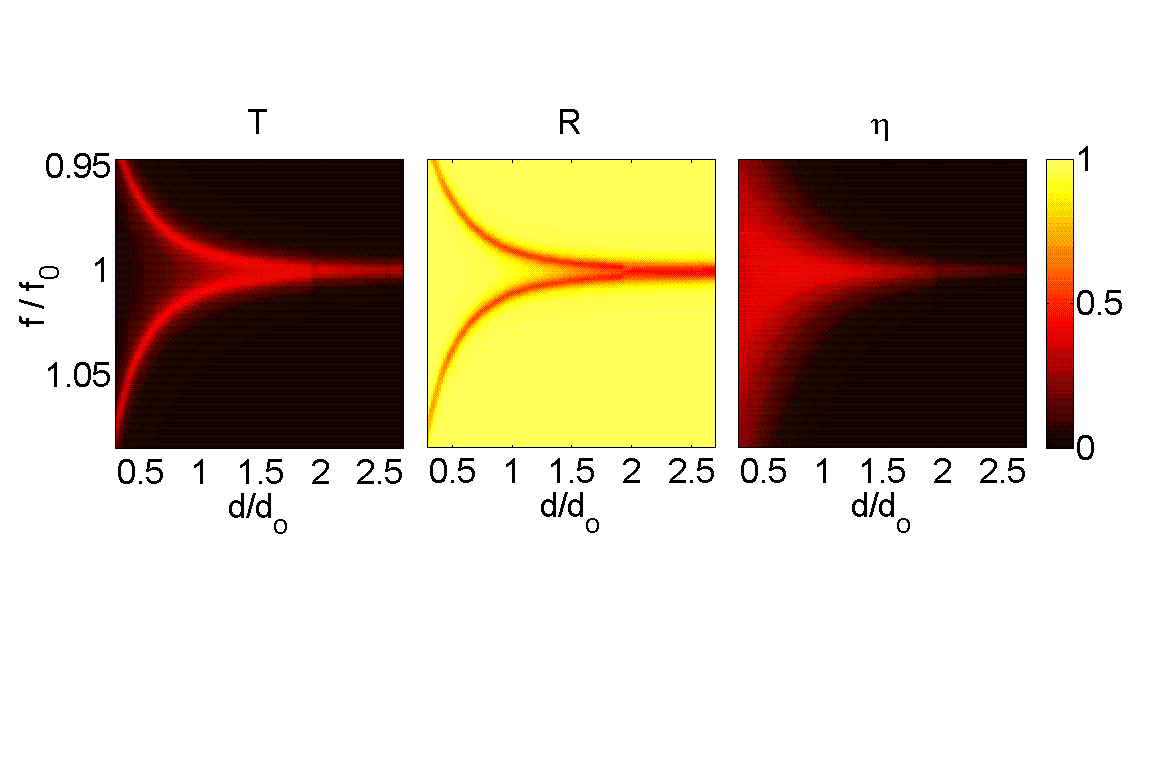}

\caption{Coupling and efficiency vs. distance. In the strong coupling regime
efficiency is high and frequency independent, but transmission is
significant only at the split frequencies.}
\label{Flo:coupFig}
\end{figure}
\begin{figure}
\includegraphics[width=1\columnwidth]{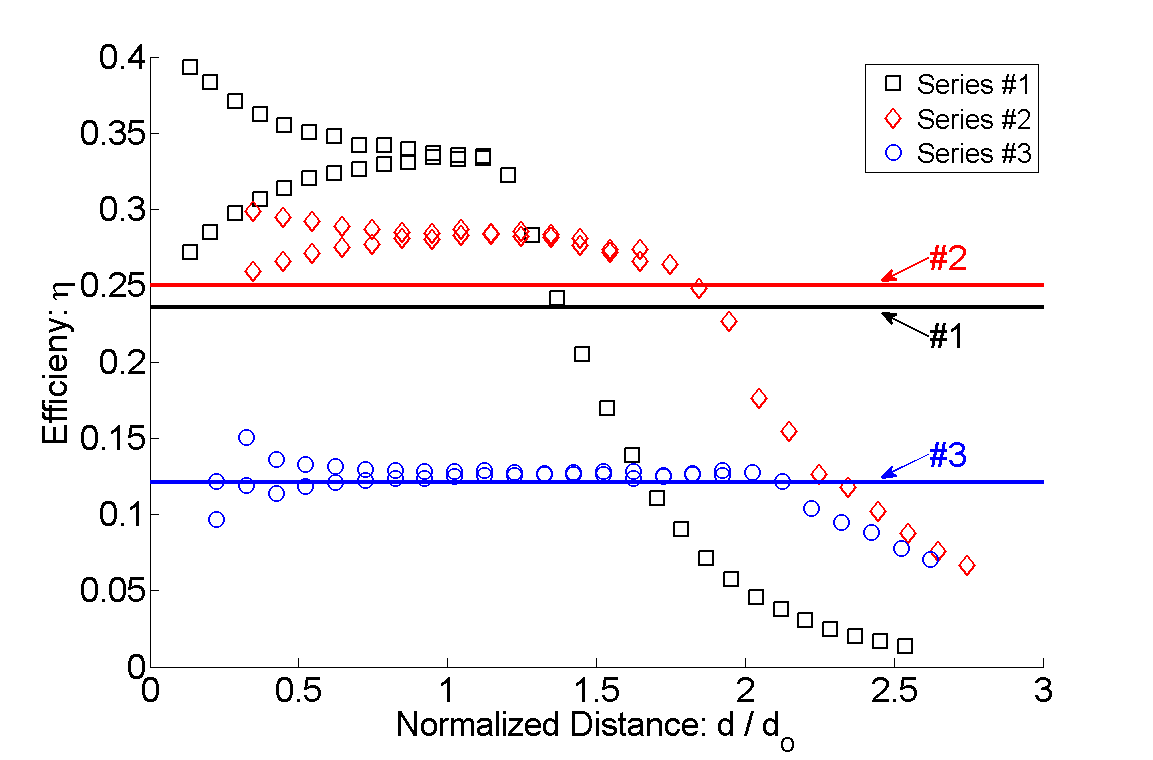}

\caption{Efficiency and range of coupling. The measured efficiency is taken
as $\eta=\frac{T^{2}}{1-R^{2}}$. The horizontal lines (labeled by
series) are those predicted by eq. \ref{eq:eff} when naively assuming
identical devices: $Q=0.5\left(Q_{1}+Q_{2}\right),k=0.5\left(k_{1}+k_{2}\right)$.}
\label{Flo:efficiency}
\end{figure}

Figs. \ref{Flo:coupFig},\ref{Flo:efficiency} detail three coupling
experiments whose parameters are given in Table \ref{Flo:coupTable}.
\begin{table}
\begin{centering}
{\scriptsize }\begin{tabular}{|c|c|c|c|c|}
\hline 
{\scriptsize Series} & {\scriptsize Device} & {\scriptsize $Q_{1},Q_{2}$} & {\scriptsize $k_{1},k_{2}$} & {\scriptsize $Q_{1}^{L},Q_{2}^{L}$}\tabularnewline
\hline
\hline 
{\scriptsize 1} & {\scriptsize 2} & {\scriptsize 139,141} & {\scriptsize 0.79,0.99} & {\scriptsize 77,71}\tabularnewline
\hline 
{\scriptsize 2} & {\scriptsize 1} & {\scriptsize 261,281} & {\scriptsize 1.11,0.89} & {\scriptsize 123,149}\tabularnewline
\hline 
{\scriptsize 3} & {\scriptsize 1} & {\scriptsize 274,261} & {\scriptsize 0.34,0.30} & {\scriptsize 205,201}\tabularnewline
\hline
\end{tabular}{\scriptsize \par}
\end{centering}

\caption{Coupled devices. {[}$Q_{m}^{L}=\frac{\omega_{0}}{2\Gamma_{m}}${]}}
{\scriptsize \label{Flo:coupTable}}
\end{table}
Fig. \ref{Flo:coupFig} shows a sample of the frequency splitting
as a function of distance, demonstrating that only at the split frequency
is power transfer both efficient and appreciable. Fig. \ref{Flo:efficiency}
compares the predicted and measured efficiency for the three measurements.
For series \#2, \#3 eq. \ref{eq:eff} gives a good prediction, with
the slight disagreement explained by the assumption of identical resonators.
For series \#1 however the efficiency considerably outstrips the estimate.
This is probably because with low Q a first order approximation no
longer properly predicts the coupling between resonators. For series
\#2, \#3 strong coupling is achieved for distances of the order $\sim2d_{o}$. 

To summarize, we have shown how a simple model of spiral inductor
resonators can be used to design coupled systems for many different
materials and size scales, by identifying crucial geometric considerations.
We have shown that proper capacitative coupling using high-Q capacitors
and increased trace width can greatly improve quality. We fabricated
integrated devices of a similar size to current handheld devices and
achieved very high Q factors and strong coupling over large distances.

\end{document}